\documentclass[useAMS,usenatbib,usegraphicx]{mn2e}

\usepackage[english]{babel}
\usepackage[utf8x]{inputenc}
\usepackage[T1]{fontenc}

\usepackage{natbib}
\usepackage{amsmath}
\usepackage{amssymb}
\usepackage{graphicx}
\usepackage[colorlinks=true, allcolors=blue]{hyperref}
\usepackage{aas_macros}
\usepackage{bm}
\usepackage{algorithm, algorithmic}
\usepackage{color}

\title[Reduced order modelling for continuous GWs]{Reduced order modelling in searches for continuous
gravitational waves -- I. Barycentering time delays}
\author[M.~Pitkin]{M.~Pitkin$^1$\thanks{E-mail: \href{mailto:matthew.pitkin@glasgow.ac.uk}{matthew.pitkin@glasgow.ac.uk}}, S.~Doolan$^1$, L.~McMenamin$^1$, K.~Wette$^2$ \\
$^1$SUPA, School of Physics and Astronomy, University of Glasgow,
University Avenue, Glasgow, G12 8QQ, UK \\
$^2$ARC Centre of Excellence for Gravitational Wave Discovery (OzGrav) and Centre for Gravitational Physics, \\
Research School of Physics and Engineering, The Australian National University, Canberra ACT 2601, Australia}

\pagerange{\pageref{firstpage}--\pageref{lastpage}} \pubyear{2018}

\begin{document}

\label{firstpage}

\maketitle

\begin{abstract}
The frequencies and phases of emission from extra-solar sources measured by Earth-bound observers are modulated by the motions of the observer with respect to the source, and through relativistic effects. These modulations depend critically on the source's sky-location. Precise knowledge of the modulations are required to coherently track the source's phase over long observations, for example, in pulsar timing, or searches for continuous gravitational waves. The modulations can be modelled as sky-location and time-dependent time delays that convert arrival times at the observer to the inertial frame of the source, which can often be the Solar system barycentre. We study the use of reduced order modelling for speeding up the calculation of this time delay for any sky-location. We find that the time delay model can be decomposed into just four basis vectors, and with these the delay for any sky-location can be reconstructed to sub-nanosecond accuracy. When compared to standard routines for time delay calculation in gravitational wave searches, using the reduced basis can lead to speed-ups of 30 times. We have also studied components of time delays for sources in binary systems. Assuming eccentricities $< 0.25$ we can reconstruct the delays to within 100s of nanoseconds, with best case speed-ups of a factor of 10, or factors of two when interpolating the basis for different orbital periods or time stamps. In long-duration phase-coherent searches for sources with sky-position uncertainties, or binary parameter uncertainties, these speed-ups could allow enhancements in their scopes without large additional computational burdens.
\end{abstract}

\begin{keywords}
gravitational waves -- methods: data analysis --  pulsars: general.
\end{keywords}

\section{Introduction}\label{sec:intro}

When examining the frequency or phase of long-duration extra-solar astronomical sources, e.g., 
pulsars, as observed using a telescope on the Earth, it is important to account for the 
frequency/phase modulation of the signal caused by the telescope's relative motion, and location 
within a gravitational potential, with respect to the source. The relative motion is caused by the 
Earth's rotation and orbital motion with respect to the Solar system barycentre (SSB), and also any 
proper motion of the source compared to the SSB. Effects of general relativistic time dilation and 
Shapiro delay must also be taken into account. If searching for weak signals, and therefore 
requiring coherent integration of long stretches of data, the precise knowledge of this modulation 
is crucial.

The coherent analysis of data over long time baselines is essential in determining the rotational
properties of pulsars. Generally, strong individual pulses are not seen, so that multiple pulses 
have to  be folded and summed, and observation periods may be short and sparsely separated, leading 
to pulse time-of-arrival measurement that are separated by a huge number of pulsar phase cycles. 
Therefore, coherently  phase matching the pulse times requires a very precise model of any extrinsic 
and intrinsic modulations of the signal. The extrinsic modulations include those caused by the 
changes between the relative inertial frames of the source and observer, such as the motion of 
detector with respect to the source described above. The ability to perform this precise phase 
matching gives a form of aperture synthesis with (for observations spanning  of order a year) a 
baseline of the Earth's orbital diameter, allowing very precise sky localisation of sources, and 
good parallax and proper motion measurement for nearby sources.

The ability to precisely localize a source is down to the fact that the specific extrinsic 
modulation will very quickly lead to decoherence of a model of the signal's phase from the true 
signal's phase as the model moves away from the true sky location. This means that for long coherent 
observations, there will be a huge number of independent phase models over the sky \citep[see, e.g., fig~14 of][which shows that, when only taking sky location into consideration, for coherent 
observation of length $T$ the number of required phase models grows
$\propto T^q$ where $q \lesssim 3$]{Wett2014:LTmPlcChASrGrvP}. The calculation of each phase model for all the independent sky 
positions can be computationally demanding, so in this paper we study using the method of reduced 
order modelling (ROM) to speed-up this computation.

ROM is a term for methods that are designed to reduce the state space 
dimensionality, or number of degrees of freedom, of a model in a way that it can be computed more 
efficiently with a  corresponding loss in accuracy. One often used ROM method is Principle Component 
Analysis in which an orthogonal basis of model vectors is constructed from a set of model vectors 
created to cover the state space of possibilities. The whole orthogonal basis can be used to 
reconstruct any model from within the original set perfectly, whilst some subset of the basis can be 
found that can reconstruct any model within the initial set to a required accuracy. As the number of 
required bases is smaller than the original state space it often allows speed-ups in calculations of 
models. In this paper we will use the method of producing an orthonormal reduced basis set described 
in section~III of \citet{PhysRevX.4.031006} \citep[also see, e.g.,][for a discussion on validation and 
enrichment methods]{2016PhRvD..94d4031S}, which we will further discuss in section~\ref{sec:rom}.

\subsection{Searches for continuous gravitational waves}

Searches for continuous sources of gravitational waves (CWs), for which the source is generally 
assumed to be a galactic neutron star with a non-zero mass quadrupole (i.e., the star has a
triaxial moment of inertia ellipsoid), assume quasi-monochromatic signals \citep[see, e.g.,][and
references therein]{2004PhRvD..69h2004A}. The signals include
the above mentioned modulations and any intrinsic frequency evolution through the inclusion
of frequency derivative terms. Due to the far smaller available mass quadruple, these sources are
intrinsically weak when compared to, for example, the final stages of the coalescence of two black 
holes  or neutron stars \citep[see, e.g.,][]{300years, 2009LRR....12....2S}. In all-sky searches for 
CWs the length of data that can be coherently analysed is generally defined by computational 
limitations based on the number of coherent phase templates required to recover signals with a
certain allowable loss in recovered power \citep[e.g.][]{1998PhRvD..57.2101B}. This compromise between coherent integration time and
computational resources has lead to the development of many heirarchical semi-coherent search methods
\citep[see, e.g.,][and references therein]{1999gr.qc.....5018S, 2000PhRvD..61h2001B, AstoEtAl2002:DAnlGrvSgSpNtSIVAS:IV, 2004PhRvD..70h2001K, 2008PhRvD..77b2001A, 2009PhRvD..79b2001A, Plet2010:PrmMSmSrCnGrvW, Wett2015:PrmMASmSrGrvPl, LIGOVirg2017:AlSrPrdGrvWvOLD}.

\subsection{Solar system barycentring}

The modulation of an extra-solar signal can, if working in terms of signal phase, be
expressed as a time modulation, e.g., for a phase evolution given by
\begin{equation}\label{eq:phase}
\phi(t) = \phi_0 + 2\upi f_0\left( t - t_0 + \Delta \tau(t) \right),
\end{equation}
where $t$ is the time of arrival of the signal at the observer, and $\phi_0$ and $f_0$ are an initial phase
and frequency at the epoch $t_0$ in a reference frame
at rest with respect to the source, the time modulation term is $\Delta\tau(t)$. Assuming, for now, that the source is at rest with respect
to the SSB, the time modulation can be expressed as a combination of terms
\begin{equation}\label{eq:deltatau}
\Delta \tau = \Delta_R + \Delta_E - \Delta_S,
\end{equation}
where $\Delta_R$ (the Roemer delay) is a geometric retardation term, $\Delta_E$ (the Einstein delay) is a
relativistic frame transformation term taking into account relativistic time dilation, and $\Delta_S$ (the Shapiro delay) is the delay due to
passing through curved space--time.
These terms are discussed in, for example, chapter 5 of \citet{LyneSmith}, while \citet{2006MNRAS.372.1549E} provide
more detailed discussion of time delays accounting for more effects with particular relevance to
pulsar observations. Here, for each of the terms we use the sign conventions given in the source 
code for the pulsar timing software {\sc tempo2}\footnote{\url{https://bitbucket.org/psrsoft/tempo2}} \citep{2006MNRAS.369..655H} and in the 
LALSuite gravitational wave software library functions \citep{LALSuite}, rather than those used in the 
equation of  \citet{2006MNRAS.372.1549E}.\footnote{In \citet{2006MNRAS.372.1549E} the equivalent of 
Equation~\ref{eq:phase} subtracts the $\Delta \tau$ term rather than adding it, and the equivalent 
of Equation~\ref{eq:deltatau} sums all the terms. These two differences mean that the Roemer delay 
and Einstein delay terms in \citet{2006MNRAS.372.1549E} have opposite signs to those used in the 
source code.} The Roemer delay is given by
\begin{equation}\label{eq:roemer}
\Delta_R = \frac{\mathbfit{r}\cdot\hat{\mathbfit{s}}}{c},
\end{equation}
where $\mathbfit{r}$ is a vector giving the position of the observer with respect to the SSB, and 
$\hat{\mathbfit{s}}$ is a unit vector pointing from the observer to the source. The Einstein delay 
\citep[see, e.g., Equations 9 \& 10 of][]{2006MNRAS.372.1549E} converts to a new time coordinate 
frame, and depends on the choice of frame you want, i.e., Barycentric Coordinate Time (TCB), in which 
the  effect of the presence of the Sun's gravitational potential is removed. The Shapiro delay (for 
which we will only consider the contribution from the Sun) is to first order given by
\begin{equation}\label{eq:shapiro}
\Delta_S \equiv \Delta_{S_{\sun}} = -\frac{2G {\rm M}_{\sun}}{c^3}\ln{\left(\mathbf{r}_{\rm se}\cdot\hat{\mathbfit{s}} + |\mathbfit{r}_{\rm se}| \right)},
\end{equation}
for waves passing around the Sun, where $\mathbf{r}_{\rm se} = \mathbf{r}_{\earth} -
\mathbf{r}_{\sun}$ is the vector from the centre of the Sun to the geocentre.\footnote{In the 
LALSuite \citep{LALSuite} functions the Shapiro delay calculation is, slightly confusingly, calculated as 
$\Delta_{S_{\sun}} = 9.852\!\times\!10^{-6} \ln{\left(1/(\mathbfit{r}_{\rm 
se}\cdot\hat{\mathbfit{s}} + |\mathbfit{r}_{\earth}|) \right)}$, which is equivalent to 
Equation~\ref{eq:shapiro} with the minus sign subsumed into the logarithm term. In {\sc tempo2} the 
$\mathbf{r}_{\rm se}$ is actually the vector from the Sun's centre and the detector,
rather than the geocentre. Using the geocentre instead leads to errors on the order of 4\,ns.} 
Unlike electromagnetic waves, gravitational waves will pass through matter, and therefore a different
term is required for a wave passing through the Sun, i.e., when
$\left|\mathbfit{r}_{\rm se}\right|^2 - \left(\mathbfit{r}_{\rm se}\cdot\hat{\mathbfit{s}}\right)^2 < R_{\sun}^2$ and 
$\mathbfit{r}_{\rm se}\cdot\hat{\mathbfit{s}} < 0$, giving
\begin{align}\label{eq:shapirothroughsun}
\Delta_{S_{\sun}} = -\frac{2G {\rm M}_{\sun}}{c^3}&\Bigg[\ln{\left(\mathbfit{r}_{\rm se}\cdot\hat{\mathbfit{s}} +
\sqrt{R_{\sun}^2 + \left(\mathbfit{r}_{\rm se}\cdot\hat{\mathbfit{s}}\right)^2} \right)} \nonumber \\
 & - 2\left(1 - \frac{\sqrt{\left|\mathbfit{r}_{\rm se}\right|^2 - \left(\mathbfit{r}_{\rm se}\cdot\hat{\mathbfit{s}}\right)^2}}{R_{\sun}} \right) \Bigg],
\end{align}
where $R_{\sun}$ is the radius of the Sun.

\begin{figure}
\includegraphics[width=\columnwidth]{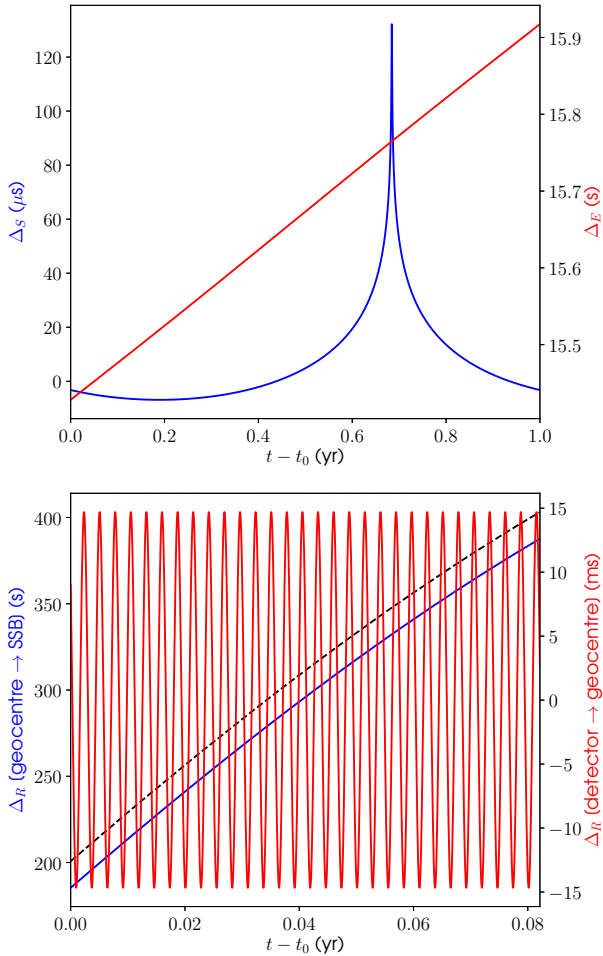}
\caption{The top panel shows the Shapiro delay (left axis, blue colour) and Einstein delay (right axis, red colour)
over one year for a particular sky position ($\alpha = 0^{\rm h}$ and $\delta = 0\degr$). The bottom panel shows
the Roemer delay from the geocentre to the SSB (left-hand axis, blue colour) and from the LIGO
Hanford Observatory to the geocentre (right-hand axis, red colour) over one month. The bottom
panel also shows the combination of all delays as the dashed black line. \label{fig:barycentre_delays}}
\end{figure}

Fig.~\ref{fig:barycentre_delays} shows the Einstein and Shapiro delays over one year, and the 
components of the Roemer delay (from the detector to the geocentre and from the geocentre to the 
SSB) for one sky location (in this case $\alpha = 0^{\rm h}$ and $\delta = 0\degr$),
and assuming a detector at the location of the LIGO Hanford Observatory.

For gravitational waves, unlike electromagnetic signals, we do not require any interstellar 
dispersion or atmospheric timing corrections, so these can be ignored.

\subsubsection{Binary system barycentring}\label{sec:bsb}

\begin{figure}
\includegraphics[width=\columnwidth]{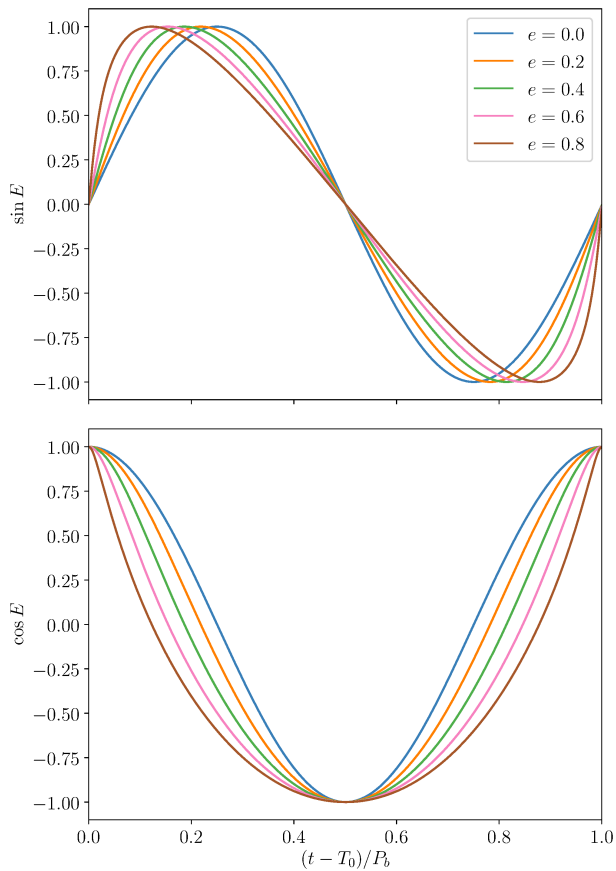}
\caption{The sine (upper panel) and cosine (lower panel) of the eccentric anomaly 
(equation~\ref{eq:eccanomaly}) for a range of eccentricities \label{fig:eccanomaly}}
\end{figure}

For sources that are in binary or multiple systems, further time delay corrections are required to 
give an inertial reference frame with respect to the source (e.g., correcting from the
binary system barycentre, BSB, to the source frame). This would add a further $\Delta_{\rm 
B}$ term to Equation~\ref{eq:deltatau}, i.e.\
\begin{equation}
\Delta\tau = \Delta_R + \Delta_E - \Delta_S - \Delta_B.
\end{equation}
$\Delta_{\rm B}$ consists of the same kind of corrections as required for the
SSB, which are nicely described in \citet{1989ApJ...345..434T}, and also in Section~2.7 of
\citet{2006MNRAS.372.1549E}. In this analysis we will only focus on binary systems and not multiple
systems, and use the model given by \citet{1976ApJ...205..580B}. From Equation~5 of
\citet{1989ApJ...345..434T} this model is given as
\begin{align}\label{eq:bt}
\Delta_B &= \left\{x \sin{\omega_0} \left(\cos{E} - e\right) + \left[x \cos{\omega_0}\sqrt{1-e^2} + \gamma\right]\sin{E}\right\} \nonumber \\
&\times \left\{1-\frac{2\pi x}{P_b}\frac{[ \cos{\omega_0}\cos{E}\sqrt{1-e^2} - \sin{\omega_0}\sin{E}]}{\left(1-e\cos{E}\right)} \right\},
\end{align}
where $x \equiv a\sin{i}/c$ is the projected semi-major axis, $\omega_0$ is the longitude of 
periastron, $P_b$ is the orbital period, $e$ is the eccentricity, $\gamma$ measures the 
gravitational redshift and time dilation (effectively conveying the Einstein delay), and $E$ is the 
eccentric anomaly defined by
\begin{equation}\label{eq:eccanomaly}
E - e\sin{E} = \frac{2\pi}{P_b}\left(t_b - T_0\right),
\end{equation}
where $T_0$ is the time of periastron passage, and $t_b \equiv t + (\Delta_R + \Delta_E - \Delta_S)$
is the barycentric time of arrival. From Equations~\ref{eq:bt} and \ref{eq:eccanomaly} it can be
seen that the eccentric anomaly is the only time varying term and this enters the binary time
delay through its sine and cosine. The eccentric anomaly for a variety of eccentricities over one
orbital period is shown in Fig.~\ref{fig:eccanomaly}. The \citet{1976ApJ...205..580B} model also 
allows time  derivatives of the orbital period, longitude of periastron and eccentricity due to 
relativistic  effects, but here we will ignore them, or assume they vary very little over the period of one orbit.

\section{Reduced order modelling}\label{sec:rom}

In the context of this paper ROM is a method for reducing the computational 
demand required to produce the model of some process, by using an approximation to that model that 
is accurate to a given level. In our case the models used are the time delays given by Equations~\ref{eq:deltatau} (without the Shapiro delay term) and \ref{eq:bt}, computed on a regular grid of
time values over one year. This relies on first producing a training set\footnote{As an explicit example of a {\it training set}, if we had a function, $f(t;L)$,
that can be evaluated at a vector of points $\mathbfit{t}$ and depends on some variable, $L$, then the
training set could consist of the set of evaluated functions $f(\mathbfit{t}; L_j)$ over a range of 
values of the variable $L$. The number of points in the vector $\mathbfit{t}$ and number of values of
$L$ defines the size of a training set array.} of models spanning some 
particular required parameter space, empirically finding an orthonormal basis model set from it that 
can match (or in our case reconstruct) the training set to a given accuracy, and creating an empirical interpolant
from it.
 In our case the empirical interpolant is just a set of times (the number of which is given by the number required orthonormal bases) in the model time series that are found to be optimal for reconstructing the model through linear operations (see Equation~\ref{eq:reconstruct}). The {\it reduced} basis and interpolant can be used to reconstruct the model (sometimes called a {\it surrogate} model) anywhere within
the bounds of the training parameter space. When the number of orthonormal bases is significantly less than the number of
points sampling the model, using the interpolant provides a speed-up in calculations, particularly 
when the model itself is a large computational burden. Here we use ROM as a catch-all term for the 
entire process of creating a reduced bases set and an empirical interpolant. The first application 
of this method, and the related {\it reduced order quadrature}, to models (and likelihoods) used in 
gravitational wave data analysis was in \citet{2011PhRvL.106v1102F}, and the algorithms are 
described in detail in \citet{PhysRevX.4.031006}. Applying the method to computationally demanding
compact binary coalescence waveform generation, involving phenomenological spinning signals in particular, has been pioneered by \citet{2014CQGra..31s5010P}
and \citet{2016PhRvD..94d4031S}.

The method used to generate the orthonormal bases from the training set uses a greedy algorithm (it {\it eats} the biggest
thing first) based on the
Gram--Schmidt process, in particular the Iterative Modified Gram--Schmidt (IMGS) algorithm \citep{Hoffmann1989}. This
is implemented in the open source {\sc greedycpp} code (Antil, Chen \& Field, in preparation), which we have modified for 
our purposes.\footnote{\url{https://github.com/mattpitkin/greedycpp/releases/tag/redordbar-v0.1}}
The {\it basic} idea (IMGS has some minor complications) for producing the basis relies on the following steps: pick a model vector from a
normalized copy of the training set, and use this as the first
basis vector; calculate the projection coefficients of the current basis (just the first
basis vector on the first pass) on to all other vectors in the training set
(by taking their dot products) and mutliply the basis by these coefficients; get the residual between these projections and the training set, and
calculate the {\it projection error} (the dot product of each residual vector with itself)
for each residual; find the residual with the maximum projection error, normalize it, and take this
as a new basis vector; expand the basis set by adding this new basis to it; this process is repeated using the expanded basis until the maximum projection error is
below some pre-defined stopping criterion, or tolerance, at which point no more bases are added.
 The general 
method is shown as pseudo-code in appendix~A of \citet{PhysRevX.4.031006}, in which the stopping 
criterion for adding new orthonormal bases to the reduced set is based on the maximum residual 
projected overlap between the current reduced basis set and the training set. As the reduced basis 
in the case of \citet{PhysRevX.4.031006} is being used to then implement {\it reduced order 
quadrature}, in which dot products of models and data are the desired output, this stopping 
criterion is sensible. However, for our analysis we are interested in the actual residuals between 
interpolated models and the training set (the errors in the time delay), and not residuals on 
projections. Therefore, our modification changes the addition of new orthonormal bases and the
stopping criterion to work with absolute residuals. Suppose our model, $h(t;\bm{\mathit{\lambda}})$, 
is defined by some vector of parameters $\bm{\mathit{\lambda}}$, and we have a training set of $M$ 
parameter values $\mathcal{T} =
\{\bm{\mathit{\lambda}}\}_{i=1}^M$ at which the model is evaluated, $\mathcal{B} = 
\{h(t;\bm{\mathit{\lambda}}_i)\}_{i=1}^M$. Our modification of the algorithm shown in \citet{PhysRevX.4.031006} is given in
Algorithm~\ref{alg:greedy}, which relies on the empirical interpolation method given in appendix~B 
of \citet{PhysRevX.4.031006}. It should be noted that while the algorithm in \citeauthor{PhysRevX.4.031006}
starts with a normalized set of training models, our method does not as it requires the absolute 
residuals.

\begin{algorithm}
\caption{A greedy algorithm for creating a reduced basis, adapted from \citet[Appendix~A]{PhysRevX.4.031006}.}
\label{alg:greedy}
\begin{algorithmic}[1]
\STATE $i=0$ and $\sigma_0 = \infty$
\STATE choose (arbitrary) seed $\bm{\mathit{\lambda}} \in \mathcal{T}$: e.g., $\bm{\mathit{\Lambda}}_1$, $\mathbfit{e}_1 = h(t;\bm{\mathit{\Lambda}}_1)$
\STATE set initial reduced basis, $\mathbfss{R} = \{\mathbfit{e}_1/|\mathbfit{e}_1|\}$
\WHILE{$\sigma_i > \varepsilon$}
  \STATE $i = i+1$
  \STATE build interpolant $\mathcal{I}_i$ from $\mathbfss{R}$
  \STATE $j=0$
  \FOR{$j < M$}
    \STATE $\mathbfit{h}_j = \mathcal{B}_j$
    \STATE create approximant $\bar{\mathbfit{h}}_j$ to $\mathbfit{h}_j$ using $\mathcal{I}_i$
    \STATE get residual error $r_j = {\rm max}|\mathbfit{h}_j - \bar{\mathbfit{h}}_j|$
    \STATE $j = j+1$
  \ENDFOR
  \STATE get index with max.\ residual error, $k = {\rm argmax}|r|$
  \STATE $\sigma_i = r_k$
  \STATE $\bm{\mathit{\Lambda}}_{i+1} = \mathcal{T}_k$
  \STATE $h = \mathcal{B}_k/|\mathcal{B}_k|$ (normalise model)
  \STATE generate $e_{i+1}$ from $h$ and RB (Gram-Schmidt)
  \STATE $e_{i+1} = e_{i+1}/|e_{i+1}|$ (normalise new basis)
  \STATE $\mathbfss{R} = \mathbfss{R} \cup e_{i+1}$ (add to reduced basis)
\ENDWHILE
\end{algorithmic}
\end{algorithm}

The output of Algorithm~\ref{alg:greedy} is a reduced basis and the set of greedy points, 
$\{\bm{\mathit{\Lambda}}_i\}_{i=1}^m$, where $m$ is the number of bases
from which the reduced basis was constructed. If the stopping criterion is not reached until all training points are
used, or $m \geq N$, where $N$ is the length of each training model, then it suggests that the training set is largely 
orthogonal already. If, for example, the training models have been evaluated at a set of $N$ time samples $t_i$, then following 
appendix~B of \citet{PhysRevX.4.031006} the reduced basis products can then be used to empirically find the $m$ values of
$t$, $\mathbfit{t}$, that can generate an optimal $m\times m$ interpolation matrix from the reduced basis
$\mathbfss{S} = \{h(\mathbfit{t},\bm{\mathit{\Lambda}}_i)\}_{i=1}^m$.

An approximant of the full model at a new point $\bm{\mathit{\lambda}}$ can then be calculated using simple linear algebra,
as it is just a weighted sum of the reduced bases. If the model is just evaluated at the $m$ points $\mathbfit{t}$,
$\mathbfit{h}' = h(\mathbfit{t};\bm{\mathit{\lambda}})$, then we can calculate the required weighting coefficients,
$\mathbfit{C}$, for combining the bases via
\begin{align}
\mathbfit{h}' &= \mathbfit{C}\mathbfss{S} \nonumber \\
\mathbfit{C} & = \mathbfit{h}'\mathbfss{S}^{-1}.
\end{align}
To reconstruct the full approximated model just requires summing the full reduced basis with appropriate weightings
\begin{align}\label{eq:reconstruct}
h(t; \bm{\mathit{\lambda}}) &= \mathbfit{C}\mathbfss{R} \nonumber \\
 & = \mathbfit{h}' (\mathbfss{S}^{-1} \mathbfss{R}).
\end{align}
The inversion of $\mathbfss{S}$ just needs to be performed once and stored, meaning the required 
operations are trivial. Later on we will refer to $\mathbfss{V} = \mathbfss{S}^{-1} \mathbfss{R}$,
which again is an operation that just needs to be performed once.

\subsection{Reduced order modelling for the SSB}\label{sec:romssb}

For this work we want to compute the value of $\Delta \tau$ in Equation~\ref{eq:deltatau} at a set of discrete
times. For a given gravitational wave detector, and a fixed time span, the computation of $\Delta \tau$ only
depends on the source's sky location. In this work we will assume that sources are far enough away that parallax
can be ignored, although this could be incorporated in the future. We therefore can create a basis training set using
parameters distributed randomly over the sky sphere, which is achieved by drawing
points uniformly in right ascension and uniformly in the sine of declination. 

Here we will take our baseline as requiring the barycentring time delay to be calculated over one year. To
create our training set we draw 5\,000 training points in right ascension and declination as described above,
and for each pair we calculate $\Delta \tau$ over one year or 365.25\,d (arbitrarily starting at 2017 January 1,
00:00:00 UTC, or a GPS time of 1\,167\,264\,018) in 60-s steps. The choices of number of training points and time-step size have been partly guided by computational memory constraints for storing the training set.  The time delays are calculated using the JPL DE405 Solar 
system ephemeris positions, velocities, and accelerations for the Sun and Earth/Moon system \citep{Standish}, and the TCB time
coordinate system. Here we have assumed signals arriving at the LIGO Hanford Observatory (H1), but we expect any Earth-bound
gravitational wave detector (or indeed any position on the Earth's surface) to produce very similar results. For our $\Delta \tau$ generation we actually do not include
the Shapiro delay term. The Shapiro delay term consists of cusps (see Fig.~\ref{fig:barycentre_delays}), with the cusp being at different points
in the time series for different sky positions (relating to when the Sun is between the Earth and the source).
Including Shapiro delay therefore makes it very difficult to produce orthogonal bases across the whole sky. However,
the $\mathbfit{r}_{\rm se}$ term in Equation~\ref{eq:shapiro} is sky position independent and therefore only needs
calculating once over the range of time steps, meaning that the computational burden for determining 
the Shapiro delay is already low.

\begin{figure*}
\includegraphics[width=1.0\textwidth]{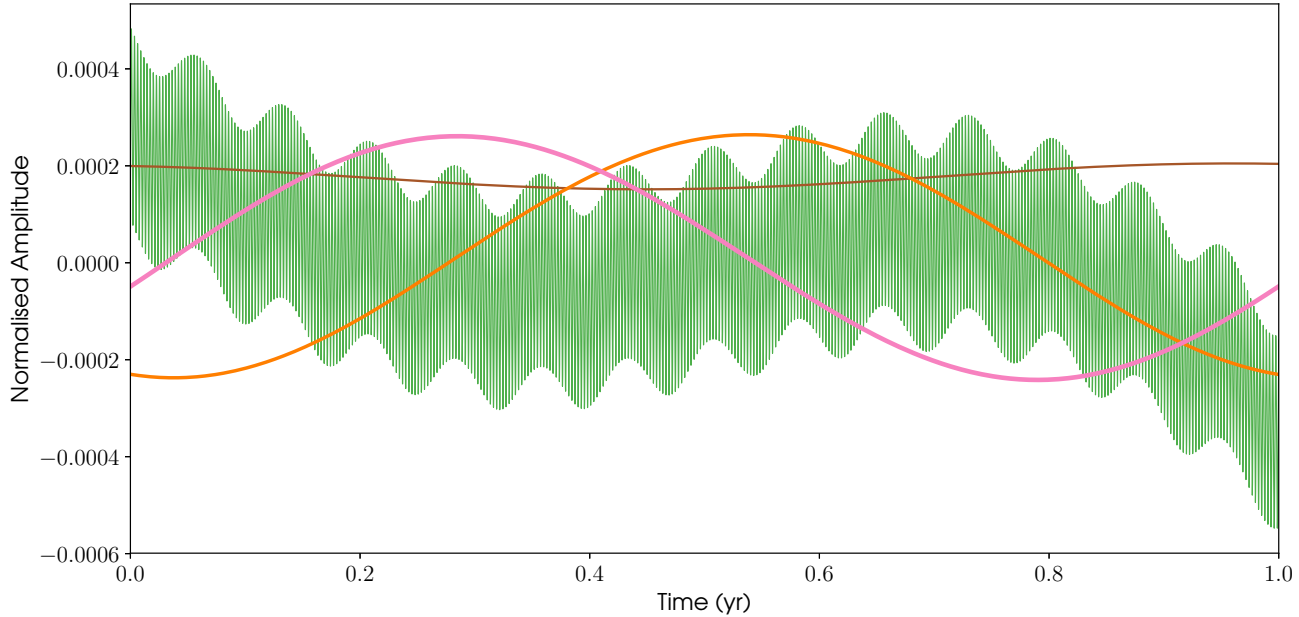}
\caption{The four orthonormal bases built from the training set covering the sky. \label{fig:reducedbases}}
\end{figure*}

We pass the 5\,000 sample training set to the modified {\sc greedycpp} code, which applies Algorithm~\ref{alg:greedy}
with the stopping criterion on adding more bases being a maximum time residual of 0.1\,s. This stopping criterion
looks surprisingly large; however, it was found to lead to a basis that actually produces far smaller time residuals
(see section~\ref{sec:accuracy}) well within the required accuracy of pulsar phase templates for gravitational wave searches.\footnote{For a
1\,kHz signal, a timing error of 100\,ns would lead to a phase error of 
$2\pi\!\times\!1000\!\times\!10^{-7} \approx 6.3\!\times\!10^{-4} {\rm rad}$, or an amplitude 
mismatch of $\sim 1-\cos{\left(6.3\!\times\!10^{-4}\right)} \approx 2\!\times\!10^{-7}$. The timing 
codes in LALSuite \citep{LALSuite} use approximations that mean they are not accurate to the same 
nanosecond precision as those in {\sc tempo2} \citep{2006MNRAS.372.1549E}, and discrepancies are 
probably on the order of a few tens of nanoseconds.}\textsuperscript{,}\footnote{Using a smaller 
tolerance, even just 0.01\,s, leads to the code including an additional unrequired fifth basis vector that appears to consist of numerical noise.} 
We find that only four orthonormal bases are required to reconstruct the training set. These bases
are shown in Fig.~\ref{fig:reducedbases}. It can be seen that there are three bases that show dominant yearly
periodicities, and one that captures yearly, monthly and daily periodicities.
As described in \citet{2016PhRvD..94d4031S} we have implemented several validation and enrichment steps to confirm that the four bases do not leave
gaps in the sky. For each validation we generate 2\,000 new randomly distributed training points to give a validation set; the reduced basis is used to create an 
interpolant to recover each model of the validation set, and any that fail the tolerance test get added to enrich the original training set. The reduced basis can 
then be rebuilt from the enriched training set. We find that the four originally recovered bases contain no gaps and no enrichment is required. In total 36\,000 
additional sky locations were tested and all could be reconstructed within the required tolerance
(see Fig.~\ref{fig:validation} and discussions in Section~\ref{sec:ssbaccuracy}).

\subsection{Reduced order modelling for the BSB}\label{sec:rombsb}

There is also potential to use ROM for the binary system barycentring time delay calculation. As discussed in
section~\ref{sec:bsb} the computational load for this delay is from the calculation of the eccentric anomaly, given by
Equation~\ref{eq:eccanomaly}. For the binary orbital delay, given in equation~\ref{eq:bt}, the eccentric
anomaly appears through its sine and cosine. So, we construct two ROMs, one for the sine term and one for 
the cosine term. For a fixed orbital period the eccentric anomaly depends on two variables,
the eccentricity, $e$, and the time of periastron passage, $T_0$. In producing a ROM we
only need to calculate the eccentric anomaly over one orbital period, as the same basis can subsequently
be used repeatedly for many orbits. We also only need to produce the ROM for one orbital period at a fixed
set of time stamps. For different orbital periods, $P_b$, the time stamps used to generate the ROM can
be scaled by $P_b^{\rm ROM}/P_b$ (where $P_b^{\rm ROM}$ is the fixed period used in calculating the ROM),
and the eccentric anomaly components reconstructed from the basis can be interpolated on to the actual
data time stamps.

\begin{figure}
\includegraphics[width=\columnwidth]{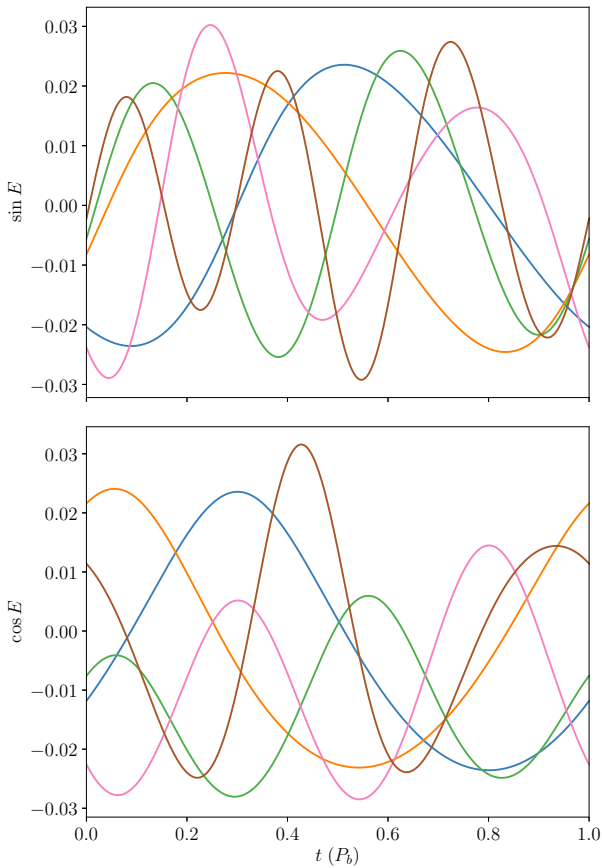}
\caption{The first five basis vectors of the sine and cosine of the eccentric anomaly for a 
maximum eccentricity of 0.25. \label{fig:eccanomalybasis}}
\end{figure}

As an example, here we constructed a ROM using a 5\,000 point training set covering a range of eccentricities
from zero to 0.25.\footnote{An eccentricity of 0.25 spans 90\% of the
measured ellipticities for binary systems containing pulsars given in v.\ 1.56 of the ATNF Pulsar Catalogue \citep{2005AJ....129.1993M}, assuming
those with ellipticities not listed in the catalogue have small values.} The eccentricities are drawn
uniformly over this range. We, somewhat arbitrarily, chose to fix the orbital period to 3\,600 s, and
therefore have training points in $T_0$ sampled uniformly between 0 and 3\,600 s. Each model ($\sin{E}$ 
and $\cos{E}$) in the training set is calculated at time stamps that span one orbital period and spaced at
one second intervals (i.e.\ there are 3\,601 time points). To produce the ROMs we set a tolerance for the
maximum timing error of 10\,ns at which to stop adding more bases.\footnote{We note that a timing 
error of 10\,ns will not necessarily be reflected in the full time delay, because we are working
with $\sin{E}$ and $\cos{E}$ rather than directly in $\Delta_B$. As such, the error on 
$\sin{E}/\cos{E}$ can get propagated into $\Delta_B$ with an increase of a factor of $\sim 
a\sin{i}/c$, where $a\sin{i}/c \sim 100$\,s is a not unreasonable value for binary orbits.} The 
first five bases for $\sin{E}$ and $\cos{E}$, given as a function of orbital period, are shown in 
Fig.~\ref{fig:eccanomalybasis}. Unlike for the Solar system barycentring ROM we are treating this as a toy problem and we have therefore not gone to the lengths of performing 
any validation or enrichment steps in the basis generation, although we do not expect there to be
significant gaps in the basis.

\begin{figure}
\includegraphics[width=\columnwidth]{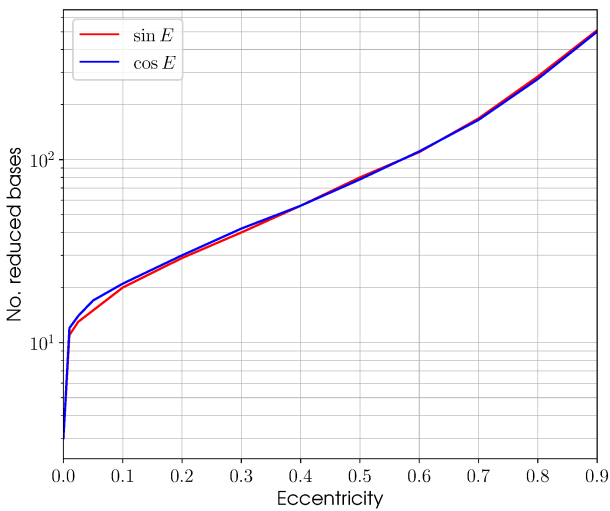}
\caption{The number of reduced bases required for reconstructing the sine and cosine of the eccentric
anomaly as a function of eccentricity. \label{fig:ecc_numbases}}
\end{figure}

We find that for an eccentricity of 0.25 the residual tolerance requires 34 bases for both the 
$\sin{E}$ and $\cos{E}$ terms, which can be compared to the 3\,601 time points used. 
Fig.~\ref{fig:ecc_numbases} also shows how the number of required bases varies with 
eccentricity. It can be seen that for fully circular orbits only three bases are required for both 
the sine and cosine components, whilst for highly eccentric orbits ($e=0.9$) over 500 bases are 
required. It is also interesting to note that for ellipticities $\lesssim 0.5$ the cosine term 
always requires bases greater than, or equal to, the number of bases for the sine term, while at larger values 
the sine term generally requires more bases.

\section{Timing and accuracy}\label{sec:accuracy}

The reason for producing the reduced bases and empirical interpolants is to provide a computational
speed-up in calculating Equations~\ref{eq:deltatau} and \ref{eq:bt} without any impact on the 
required precision. Here we provide timings of these calculations when evaluating the functions 
explicitly at all required time-steps and compare them to the timings when using the reduced basis 
and empirical interpolants.\footnote{All timings have been performed on an Intel Core i5-4570 CPU @ 
3.2\,GHz. Any {\tt C}-language code used has been compiled using the GNU Compiler Collection (gcc)
version 5.4.0, using the {\tt -O3} optimization flag.} We also assess the accuracy of the reconstructed
time delays by looking at the residual differences between the full evaluations and interpolated versions.

We should note that the production of the ROM does require some overhead: of the order of 10 min for 
the Solar system barycentring time delays, and tens of seconds for the eccentric anomaly. However, 
these overheads are one-off requirements compared to the huge number of times the full functions 
might be needed.

\subsection{Solar system barycentring}\label{sec:ssbaccuracy}

\begin{figure}
\includegraphics[width=1.0\columnwidth]{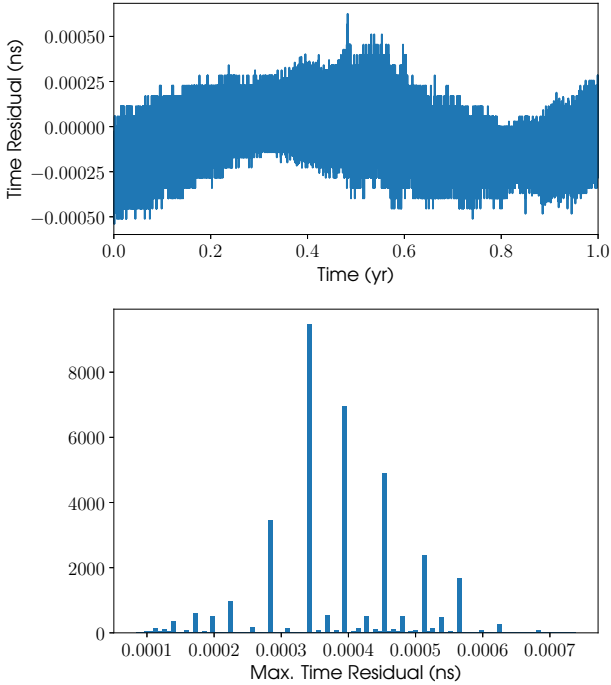}
\caption{The upper panel shows an example of the timing residuals between the explicit evaluation
of $\Delta\tau$ over one year and the reconstruction using the ROM and empirical interpolant. The
lower panel shows a histogram of the absolute maximum timing residual for reconstructions at 36\,000
sky locations.
\label{fig:validation}}
\end{figure}

Above we found that the model given by Equation~\ref{eq:deltatau} (with the Shapiro delay term 
removed) can be approximated for any position on the sky using just four basis vectors. During the
validation of the reduced basis a total of 36\,000 sky points were tested, for which the basis
and empirical interpolant generated from it were used to reproduce the time delay over the full
grid of time-steps at which the basis was produced. Residual time delays calculated by subtracting
these reconstructed delays from those explicitly calculated for the given sky position have a
maximum absolute value of $7.4\!\times\!10^{-4}$\,ns. A histogram of the maximum absolute residual 
for each of the 36\,000 test points is shown in Fig.~\ref{fig:validation}, along with an example 
of the residual time series (which in this case also includes the Shapiro delay as calculated using 
equations~\ref{eq:shapiro} and \ref{eq:shapirothroughsun}, with $\mathbfit{r}_{\rm se}$ and 
$\left|\mathbfit{r}_{\rm se}\right|$ pre-computed). The accuracy achieved is far better than any 
requirements for residual phase uncertainties in a pulsar model.

We also need to show that this basis can speed-up calculations when compared to explicitly 
calculating the time delay at the same number of time-steps as used in the basis. When using the
ROM to reconstruct the time delay for a given sky position there are several things that need to be 
calculated: there is a one-off evaluation of $\mathbfss{V} = \mathbfss{S}^{-1}\mathbfss{R}$ from 
Equation~\ref{eq:reconstruct}; the time delay needs to be explicitly calculated at the four 
empirical interpolant nodes, $\mathbfit{h}'$; the dot product $\mathbfit{h}'\mathbfss{V}$; the 
Shapiro delay, using pre-computed sky-position independent vectors for $\mathbfit{r}_{\rm se}$ and 
$\left|\mathbfit{r}_{\rm se}\right|$; and, finally, the reconstructed time delay and Shapiro delay 
need to be combined. The timings for these various steps, and the ordering of the steps that must 
be re-computed for different parameters, are given in Table~\ref{tab:ssbtimings}. To explicitly 
calculate the SSB time delay at the 525\,960 time points over the year takes $\sim 
7.6\!\times\!10^{5}\,\mu$s, whilst using the ROM takes $A_1+B_1+C_1+D_1 \approx 
2.7\!\times\!10^{4}\,\mu$s (where we ignore the one-off calculations), showing a speed-up factor of 
$\lesssim 30$. The vast majority of the computation time comes from the Shapiro delay step.

\begin{table}
\centering
\caption{Computational evaluation times for aspects of the Solar system barycentring time delay}
\label{tab:ssbtimings}
\begin{tabular}{l l c}
\hline
Order & Function & Time ($\mu s$) \\
\hline
 & $\Delta\tau$ (525\,690 evaluations)$^a$ & $\sim 7.6\!\times\!10^{5}$ \\
 & $\Delta\tau$ (1 evaluation)$^a$ & $\sim 1.4$ \\
 & $\mathbfss{V} = \left(\mathbfss{S}^{-1}\mathbfss{R}\right)$$^b$ & $\sim 1.1\!\times\!10^{4}$ \\
$A_1$ & $\mathbfit{h}' = \Delta\tau^{\rm no~Shap.}$ (4 evaluations) & $\sim 5.7$ \\
$B_1$ & $(\Delta\tau)_{\rm int}^{\rm no~Shap.} = \mathbfit{h}'\mathbfss{V}$ & $\sim 1.6\!\times\!10^{3}$ \\
$C_1$ & Shapiro delay$^c$ &  $\sim 2.4\!\times\!10^{4}$ \\
$D_1$ & $(\Delta\tau)_{\rm int}^{\rm no~Shap.} -$ Shapiro delay & $\sim 820$ \\
\hline
\multicolumn{3}{l}{$^a$ Equation~\ref{eq:deltatau}} \\
\multicolumn{3}{l}{$^b$ see Equation~\ref{eq:reconstruct}} \\
\multicolumn{3}{l}{$^c$ Shapiro delay calculated using pre-computed $\mathbfit{r}_{\rm se}$ \& $\left|\mathbfit{r}_{\rm se}\right|$} \\
\end{tabular}
\end{table}

\subsubsection{Further interpolation}

The training set, and therefore the reduced basis vectors, for the SSB delay is
calculated on a time grid at 60-s intervals. However, values of the delay at time values 
between these grid points may well be required. Such values can be calculated by creating
interpolation functions for the reduced bases. When reconstructing a time delay vector we work
with the matrix $\mathbfss{V} = \mathbfss{S}^{-1}\mathbfss{R}$ from Equation~\ref{eq:reconstruct},
which in this case is $4 \times 525\,690$ in size. For each of the four rows we can create a cubic
B-spline interpolation function [using the {\tt splrep} and {\tt splev} functions from {\sc scipy}'s
\citep{scipy} {\tt interpolate} module for generating and evaluating the interpolation function,
respectively]; we can also do this for the vectors required for the Shapiro delay calculation:
the three components of $\mathbfit{r}_{\rm se}$ and $\left|\mathbfit{r}_{\rm se}\right|$.
Creation of each of these interpolation functions takes $\sim 10^5\,\mu$s each, although these are 
one-off evaluations.

\begin{figure}
\includegraphics[width=1.0\columnwidth]{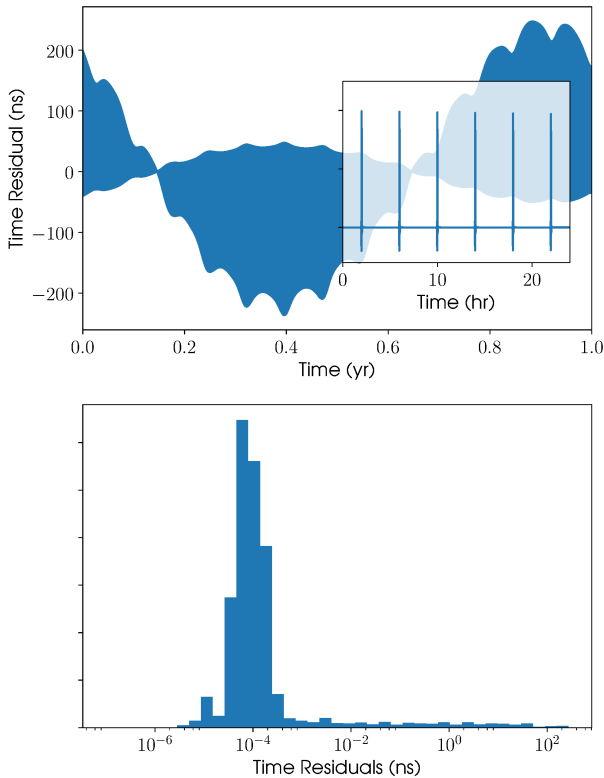}
\caption{The upper panel shows an example of the timing residuals between the explicit evaluation
of $\Delta\tau$ over one year and the reconstruction using the ROM interoplated using a cubic B-spline
at time stamps half-way between those used to create the ROM basis. The inset in the upper panel shows
a zoom of the residuals for one day. The lower panel shows a histogram of the absolute timing
residual for reconstructions at 100 sky locations.
\label{fig:interpres}}
\end{figure}

We have tested the accuracy of the interpolations by creating a new set of time-steps half-way
between those used for the basis generation, i.e.\ at 60 second intervals, but all maximally offset 
by 30 s. Each of the eight spline interpolation functions (the four rows of $\mathbfss{V}$ and the 
four vectors for the Shapiro delay calculation) was evaluated at these new time stamps, to give new 
matrices $\mathbfss{V}'$, $\left|\mathbfit{r}_{\rm se}\right|'$, and $\mathbfit{r}_{\rm se}'$. The 
evaluations each took $\sim 3.0\!\times\!10^4\,\mu$s, but again these are sky-position independent
and therefore one-off calculations, so do not affect overall speed-up of the time delay calculation. 
From these, the time delays were reconstructed as above for 100 random sky locations. Examples of the residuals between
them and explicit evaluation of the time delays at the new time points, along with a histogram of the residuals,
are shown in Fig.~\ref{fig:interpres}. In the example residuals, it would appear at first glance that the
errors are of the order of hundreds of nanoseconds, which are acceptable, but not as good as might
be hoped given the extraordinary agreement when not interpolating between time-steps shown in Fig.~\ref{fig:validation}. However,
the zoom-in of the time series also shown in Fig.~\ref{fig:interpres}, and the histograms of residuals, 
shows that the largest errors occur as spikes every four hours, whilst between these spikes the 
residuals are far smaller; we find that $\lesssim 8\%$ of times produce residuals larger than the value of
$7.4\!\times\!10^{-4}$\,ns seen above as the maximum value in Fig.~\ref{fig:validation}, $\lesssim 2.6\%$
have residuals larger than 1\,ns, and $\lesssim 1.2\%$ have residuals larger than 10\,ns.
We find that these spikes are due to the 4-h sample rate of Earth ephemeris (position, velocity, and accelations) values \citep[derived from the JPL ephemerides][]{Standish} used in the ephemeris files within LALSuite (the Sun
ephemeris has a 40-h sample rate that falls on the time bins for the Earth data); within the LALSuite barycentring routines
the extrapolation of the Earth's position and velocity between these samples leads to minor
discontinuities at the joins (half way between each sample), and the cubic spline produces an
impulse response at these joins.

The full years worth of time delay calculations will not be required for all analyses, many of
which would use shorter subsets of data. Therefore, using the full length of the reduced basis when
calculating the the time delays over the required period would be more computationally demanding 
than necessary. However, the full reduced basis is not required and the a submatrix of the reduced 
basis matrix $\mathbfss{R}$ in Equation~\ref{eq:reconstruct} can just be used that spans the 
required times values.

\subsection{Binary system barycentring}

For the binary system time delay model described in Sections~\ref{sec:bsb} and \ref{sec:rombsb} we 
found that for eccentricities of $\le 0.25$ 34 reduced bases were required for both the $\sin{E}$ 
and $\cos{E}$ terms. The evaluation times for various functions are provided in 
Table~\ref{tab:eccAeval}, where the order of operations required for the ROM interpolant (following 
its generation) are given. The time to calculate $\Delta_B$ using the ROM interpolant is 
therefore given by $\sim 2A_2 + 2B_2 + C_2 \approx 130$\,$\mu$s. Compared to the reference time of
1\,600\,$\mu$s given in the first line of Table~\ref{tab:eccAeval} this shows that using the 
interpolant can speed up the computation by $\sim 12$ times.

\begin{table}
\centering
\caption{Computational evaluation times for aspects of the binary system time delay}
\label{tab:eccAeval}
\begin{tabular}{l l c}
\hline
Order & Function & Time ($\mu s$) \\
\hline
 & $\Delta_B$ (3\,601 evaluations)$^a$ & $\sim 1\,600$ \\
 & $\Delta_B$ (1 evaluation)$^a$ & $\sim 0.5$ \\
 & $\mathbfss{V}_{\sin{E}/\cos{E}} = \left(\mathbfss{S}^{-1}\mathbfss{R}\right)$$^b$ & $\lesssim 2\,200$ \\
$A_2$ & $\mathbfit{h}' = \sin{E}, \cos{E}$ (34 evaluation) & $\sim 20$ \\
$B_2$ & $(\sin{E}/\cos{E})_{\rm int} = \mathbfit{h}'\mathbfss{V}$ & $\sim 27$ \\
$C_2$ & $\Delta_B$ (using $(\sin{E}/\cos{E})_{\rm int}$)$^c$ & $\sim 38$ \\
\hline
\multicolumn{3}{l}{$^a$ Equation~\ref{eq:bt}} \\
\multicolumn{3}{l}{$^b$ see Equation~\ref{eq:reconstruct}} \\
\multicolumn{3}{l}{$^c$ using Equation~\ref{eq:bt}, but with interpolated $\sin{E}$ and} \\
\multicolumn{3}{l}{\phantom{$^c$ }$\cos{E}$ vectors.}
\end{tabular}
\end{table}

\begin{figure}
\includegraphics[width=\columnwidth]{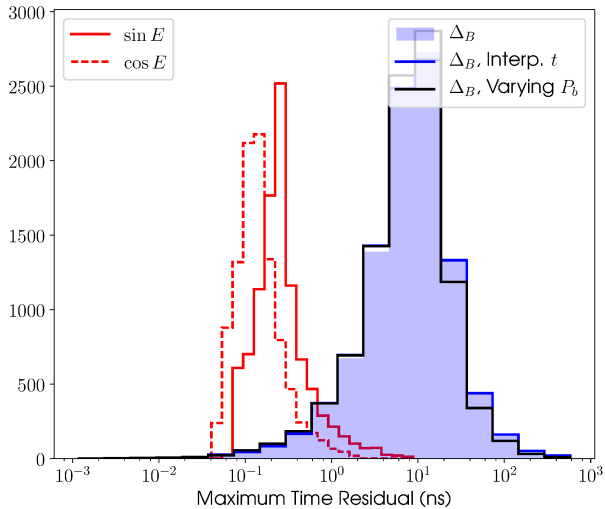}
\caption{Histograms of maximum residuals between the functions $\sin{E}$, $\cos{E}$ and 
$\Delta_B$ as explicitly evaluated and reconstructed using the ROM. \label{fig:bt_accuracy}}
\end{figure}

The accuracy (maximum residual time) with which the ROM reconstructs both the sine/cosine of 
the eccentric anomaly and the binary time delay $\Delta_B$ over one orbital period is shown in 
Fig.~\ref{fig:bt_accuracy} for 10\,000 randomly generated set of binary system 
parameters.\footnote{All of the system parameters bar the period are drawn uniformly from the
following ranges: $e \in [0, 0.25]$,  $a\sin{i}/c \in [0,100]$\,s, $T_0 \in [0, P_b]$\,s,
$\omega_0 \in [0, 2\pi]$\,rad, and $\gamma \in [0, 0.001]$\,s. $a\sin{i}/c$ values of less than 
100\,s cover 97\% pulsars in binary systems with eccentricities less than 0.25
based on v.\ 1.56 of the ATNF Pulsar Catalogue \citep{2005AJ....129.1993M}. The value of the period 
is discussed in the main text.} The red solid and dashed histograms show that the reconstructed 
$\sin{E}$ and $\cos{E}$ values are always within the tolerance of 10\,ns used when producing the 
reduced basis. The histograms shown as the shaded area and the solid blue line show cases where the
binary period for all systems has been held fixed at the 1-h value used in the training set 
generation. For the histogram shown as the solid blue line, the ROM reconstructed binary time delays 
have been subsequently interpolated (using a cubic B-spline interpolator) on to time-steps half-way 
between those used for the ROM generation, and compared to the evaluation of the full time delay at 
those new time-steps. The histogram represented by the black solid line shows the case when the 
period has been drawn uniformly between 10 mins and 10 h, for which the interpolant time-steps 
have been rescaled by $(P_b/1\,{\rm hour})$, and again a cubic B-spline interpolator used to
re-sample to the required time-steps. The accuracy of $\Delta_B$ is not able to match that for 
$\sin{E}$ and $\cos{E}$ alone, as the errors on these propagate into $\Delta_B$ with a potential 
increased by a factor on the order of the value of $a\sin{i}/c$, which leads to larger errors. 
However, we find that for all cases (given that our range of $a\sin{i}/c$ values are restricted to $< 
100$\,s)  less than 2\% of our systems lead to errors on $\Delta_B$ of greater than 100\,ns, with 
maximum errors of $\sim 500$\,ns.

As mentioned above, it may be required to evaluate the function $\Delta_B$ at time-steps that
are not the same as those used in the training set generation, or for different binary
periods. This means that further interpolation is required, which could reduce the speed-up
that the ROM interpolation gives. If, as for the SSB case, we use a cubic B-spline, then the 
interpolation function generation using 3\,601 time points takes $\sim 500\,\mu{\rm s}$ and 
evaluation, again at 3\,601 new time points, takes $\sim 270\,\mu$s. So, if interpolation to 
new times is required the total time is (again taking values from Table~\ref{tab:eccAeval}) $\sim 130 
+ 500 + 270 \approx 900\,\mu{\rm s}$, which only shows a speed-up of just under two times.\footnote{The
overhead of calling the {\sc scipy} interpolation functions is not known, so a purely
{\tt C}-based code {\it may} show a better speed-up factor.} Unlike 
for the SSB case, if the binary period is a parameter that is changing, then we cannot produce 
one-off interpolation functions for each component of the $\mathbfss{V} = 
\mathbfss{S}^{-1}\mathbfss{R}$ matrix, as the interpolation time-steps have to be scaled by the period.

Performing a simpler linear interpolation only takes $\sim 100\,\mu{\rm s}$, but produces a mean 
error in the time delay of $\sim 10\,000$\,ns with a maximum around 1\,ms.

\section{Application}

It is useful to estimate how the speed-up demonstrated above could be applied in real searches for
continuous gravitational wave signals.  The time $\tau$ taken to perform coherent matched-filtering
of a single continuous wave signal template can be modelled as \citep{PrixFstatTiming}:
\begin{equation}\label{eq:timingmodel}
\tau = \tau_{\text{core}} + b \tau_{\text{bary}}
\end{equation}
when $\tau_{\text{core}}$ is the time taken to perform the core filtering operations,
$\tau_{\text{bary}}$ is the time taken to perform solar system and/or binary system barycentering.
It is usual to buffer the barycentered time series if the sky position and/or binary system
parameters have not changed from the previously-analysed signal template; conversely when the sky
position and/or binary system parameters do change, the barycentered time series must be recomputed
for the new parameters. The fraction of signal templates where barycentering must be re-performed is
denoted $b$.  The value of $b$ largely depends on the design of the search algorithm and the
parameter-space being searched.  For search algorithms that do not use a search grid but rather
compute templates at randomly chosen parameters \citep[e.g.][]{2013PhRvD..87h4057S,2014PhRvD..89l4030S,AshtonPyFstat,AshtonPrix2017} 
we have $b = 1$.

The speed-up of such a search $x_{\text{search}}$ due of the use of a ROM may be
quantified using Eqn.~\eqref{eq:timingmodel} as
\begin{equation}\label{eq:speedup}
\begin{split}
x_{\text{search}} &= \frac{ \tau_{\text{core}} + \tau_{\text{bary}} }{ \tau_{\text{core}} + \tau_{\text{bary}} / x_{\text{ROM}} } \\
&= \frac{ 1 + x^{\text{bary}}_{\text{core}} }{ 1 + x^{\text{bary}}_{\text{core}} / x_{\text{ROM}} }
\end{split}
\end{equation}
where $x_{\text{ROM}}$ is the speed-up from the use of a ROM, and $x^{\text{bary}}_{\text{core}}$ is
the fraction of time spent computing the (non-ROM) barycentering, relative to performing the core
filtering.  Representative values of $x^{\text{bary}}_{\text{core}}$ are $\sim 17$ for sky
demodulation and $\sim 23$ for binary demodulation.\footnote{This assumes matched filtering is
  performed using the demodulation algorithm of \citet{WillSchu2000:EfMFlAlDCntGrvWS}; the FFT-based
  resampling algorithm of \citet{JaraEtAl1998:DAnGrvSgSpNSSDtc} is only efficient when searching over
  a wide frequency range.}  Given a speed-up of $x_{\text{ROM}} \sim 30$ or $12$ from the use of a
ROM for sky or binary demodulation, potential search speed-ups are $x_{\text{search}} \sim 11$ or $8$
respectively.

\section{Conclusions}

In this paper we have aimed to show that ROM can be used as a way to approximate
the time delays required to transform a signal received at an observatory on Earth to the inertial
frame of the Solar system (or binary system) barycentre. For the Solar system this transformation is
sky-position dependent, and if requiring coherent integration of signals over long periods its
recalculation can become a computational burden. In particular, this could be an issue for some large sky area
searches for continuous gravitational wave sources, or the long-coherent time follow-up of candidates from such searches
\citep{2013PhRvD..87h4057S,2014PhRvD..89l4030S,AshtonPyFstat,AshtonPrix2017}. We have shown that the Solar system barycentring
time delay function can be very well approximated using just four basis vectors, when excluding the
Shapiro delay. Using this reduced basis can significantly speed up the calculation of the time delays
by up to a factor of 30, even when adding on the computation of the Shapiro delay term. In general the 
reconstructed time delays are accurate to sub-nanosecond precision when compared to the full 
calculation. If time delays needs to be calculated at time stamps not used for the reduced basis production,
then additional interpolation of the basis vectors can be used. In this case it, is found that for a few 
percent of the samples the reproductions are accurate only to within $\sim 100$\,ns. This larger residual has 
been found to be a feature of the sample rate of the Solar system ephemeris files within LALSuite \citep{LALSuite}.

In cases where SSB time delay calculations are a bottleneck in analyses, for example, if having to 
search over a large number of sky positions in long data sets, this will reduce the computational 
burden and may allow larger parameter spaces to be searched. The barycentre time delay model used in 
this work does not include all the components used in, for example, the pulsar timing software {\sc 
tempo2} \citep{2006MNRAS.372.1549E}. However, in the future it may well be straightforward to 
incorporate the {\sc tempo2} timing model into {\sc greedycpp} (Antil et al., in preparation). In the future
the timing routines in LALSuite \citep{LALSuite} could be adapted to incorporate the ROMs, as produced
using the {\sc tempo2} model, and thus ensure consistency between the two code bases.

In addition to SSB time delays, we have also looked at the calculations required for time delay in
binary systems. The main computational burden in such calculations is the eccentric anomaly. We have
shown that for eccentricities of $< 0.25$ a reduced basis of 34 vectors is required to reproduce the
sine and cosine of the eccentric anomaly to a precision of less than 10\,ns. In the simplest cases,
when not varying the binary orbital period, factors of $\sim 10$ speed-up in the time delay calculation
are found.

Outside of the initial application of this method for continuous {\it gravitational wave} sources with
Earth-bound detectors, there are other areas where it could be used. For third-generation gravitational
wave detectors (e.g., the {\it Einstein} Telescope, \citealp{ET}, or Cosmic Explorer, \citealp{2017CQGra..34d4001A}),
the signals from compact binary coalescences may be within the sensitivity bands for days, so coherent
integration will have to account for Earth rotational and orbital motion using the delays discussed here.
For future space-based gravitational wave detectors \citep[e.g., LISA][]{2013GWN.....6....4A, LISA} the majority of the expected signals
will be quasi-continuous and long-lived within the detector's sensitive band. The orbital motion of the spacecraft will need to be
accounted for when searching for signals and the ROM method applied here could be very useful.
This approach may also be useful for standard pulsar timing applications, especially in cases where the
number of time-of-arrival observations become large, and sky positions need to be incorporated into 
parameter fits. Methods that have to sample over parameter spaces such as {\sc temponest} \citep{2014MNRAS.437.3004L}
or {\sc bayesfit} \citep{2014MNRAS.440.1446V} may particularly benefit from faster model evaluations.

In a future paper we will study how ROM methods, and the related {\it reduced order quadrature},
can be used to speed-up likelihood evaluations in searches for gravitational waves from pulsars.
The method has already been applied in the search of LIGO data for signals from known pulsars in
\citet{2017ApJ...839...12A}.

\section*{Acknowledgements}

This work has benefited greatly from discussions with Rory Smith, and from many
discussions with members of the LIGO Scientific Collaboration and Virgo Collaboration,
in particular members of the continuous waves working group. The analysis has
relied on the {\sc greedycpp} software (Antil et al., in preparation) and LALSuite \citep{LALSuite}. The analysis
has also been greatly aided by the use of IPython \citep{PER-GRA:2007}, jupyter notebooks 
\citep{kluyver2016jupyter}, and Cython \citep{behnel2010cython}, and all plots have been produced 
using Matplotlib \citep{Hunter:2007,michael_droettboom_2017_248351}. 
MP is funded by the Science \& Technology Facilities Council under 
grant number ST/N005422/1. KW is supported by the Australian Research Council CE170100004. This paper carries LIGO Document Number LIGO-P1700373.

\bibliographystyle{mn2e}
\bibliography{main}

\label{lastpage}

\end{document}